\documentclass[preprint,prc,aps,showpacs,showkeys,groupedaddress,floatfix,superscriptaddress]{revtex4-1}
\usepackage{epsfig}
\usepackage{dcolumn}
\usepackage{bm}
\usepackage[utf8]{inputenc}
\usepackage{graphics}
\usepackage{graphicx}
\usepackage{epsfig}
\usepackage{amssymb}
\usepackage{amsmath}
\usepackage{color}

\begin{document}
\title{A statistical mechanical analysis on the bound state solution of an energy-dependent deformed Hulth\'en potential energy}
\author{B.C. L\"{u}tf\"{u}o\u{g}lu}
\affiliation{Department of Physics, Faculty of Science, Akdeniz University, 07058
Antalya, Turkey}
\affiliation{Department of Physics, Faculty of Science, University of Hradec Kr\'{a}lov\'{e}, Rokitansk\'{e}ho 62, 500\,03 Hradec Kr\'{a}lov\'{e}, Czechia}

\author{A.N Ikot}
\affiliation{Department of Physics, Theoretical Physics Group, University of Port Harcourt, Choba, Port Harcourt, Nigeria.}
\author{U.S. Okorie}
\affiliation{Department of Physics, Theoretical Physics Group, University of Port Harcourt, Choba, Port Harcourt, Nigeria.}
\affiliation{Department of Physics, Akwa Ibom State University, Ikot Akpaden, P.M.B. 1167, Uyo, Nigeria.}

\author{A.T. Ngiangia}

\affiliation{Department of Physics, Theoretical Physics Group, University of Port Harcourt, Choba, Port Harcourt, Nigeria.}

\date{\today}
\begin{abstract}
In this article, we investigate the bound state solution of the Klein Gordon equation under mixed vector and scalar coupling of an energy-dependent deformed Hulth\'en potential in D-dimensions.  We obtain a transcendental equation after we impose the boundary conditions. We calculate energy spectra in four different limits and in arbitrary dimension via the Newton-Raphson method.  Then, we use a statistical method, namely canonical partition function, and discuss the thermodynamic properties of the system in a comprehensive way.   We find out that some of the thermodynamic properties overlap with each other, some of them do not.
\end{abstract}
\maketitle 

\section{Introduction}
One of the major investigation areas in either relativistic or non-relativistic quantum mechanics is to obtain a solution of potential energies  \cite{Schiff_Book, Landau_Book, Flugge_Book, Greiner_Book}. This intense interest is based on the fact that the exact solution of the wave function has all the necessary information to define the physical system. Unfortunately, only a few numbers of potential energies have exact solutions. For instance infinite well, finite well or barrier, Coulomb potential, and harmonic oscillator. Beside these analytic solutions,  semi exact solutions in case of $l=0$, or approximate solutions in case of $l\neq 0$ are investigated comprehensively in  many other potential energies such as Morse \cite{Morse_1929}, Eckart \cite{Eckart_1930}, Rosen-Morse (RM) \cite{Rosen_Morse_1932}, Manning-Rosen (MR) \cite{Manning_Rosen_1933}, P\"oschl-Teller (PT) \cite{Poschl_Teller_1933}, Yukawa \cite{Yukawa_1935},  Hylleraas \cite{Hylleraas_1935}, Hulth\'en \cite{Hulthen_1942}, Woods-Saxon (WS) \cite{Woods_Saxon_1954}, etc.

The Klein–Gordon (KG) equation is one of the fundamental relativistic wave equation that  describes the motion of spin zero particles \cite{Klein_1926}. Remarkable efforts have been executed to examine the solutions of the KG equation with a various number of  potential energies. Yi \emph{et al.} employed RM type vector and scalar potential energies to obtain the s-wave bound state energy spectra \cite{Yi_et_al_2004}. Villalba \emph{et al.} examined the bound state solution of a spatially one-dimensional cusp potential energy in the KG equation \cite{Villalba_et_al_2006}. Olgar \emph{et al.} employed a supersymmetric technique to obtain a bound state solution of the s-wave KG equation with equal scalar and vector Eckart type potential energy \cite{Olgar_et_al_2006}. Only two years later, they applied the asymptotic interaction method (AIM), which is originally introduced by Ciftci \emph{et al.} \cite{Ciftci_et_al_2003},  to calculate an energy spectrum of the s-wave KG equation with the mixed scalar and vector generalized Hulthén potential in one dimension \cite{Olgar_et_al_2008}. Then, he used AIM to investigate bound state solution of three different potential energies, namely linear, Morse and Kratzer, in the KG equation \cite{Olgar_2008}. In 2010, Xu \emph{et al.} studied the bound state solution of the KG  equation with mixed vector and scalar PT potential energy with a non zero angular momentum parameter \cite{Xu_et_al_2010}. Ikot \emph{et al.} obtained an exact solution of the Hylleraas potential energy in the KG equation \cite{Ikot_et_al_2012}. Jia \emph{et al.} examined the bound state solution of the KG equation with an improved version of the MR potential energy \cite{Jia_et_al_2013}. Hou \emph{et al.} studied the bound state solution of the s-wave KG equation with vector and scalar WS potential energy \cite{Hou_et_al_1999}. Rojas \emph{et al.} used the vector  WS barrier in the KG equation and presented the continuum state solution \cite{Rojas_et_al_2005}. Later, Hassanabadi extended that study with an addition of scalar WS potential energy term \cite{Hassanabadi_et_al_2013}. Arda \emph{et al.} employed Nikiforov-Uvarov (NU) and studied the modified WS potential energy with position dependent mass in the KG equation in three dimensions \cite{Arda_et_al_2009}. Badalov \emph{et al.} used NU and Pekeris approximation to study any $l$ state of the KG equation \cite{Badalov_et_al_2010}. Bayrak \emph{et al.} investigated the generalized WS potential energy in the KG equation for zero \cite{Bayrak_et_al_2015D} and non-zero \cite{Bayrak_et_al_2015E} values of the angular momentum parameter.  One of the authors of this manuscript, L\"utf\"uo\u{g}lu,  with his collaborators examined the  mixed vector and scalar generalized symmetric WS potential energies for the scattering case in the KG equation first under the equal magnitudes and signs (EMES), and then, in the equal magnitudes and opposite signs (EMOS)  \cite{Lutfuoglu_et_al_2018_LK}. Later, he investigated the same problem in the bound state case \cite{Lutfuoglu_et_al_2018}. Beside these studies, multi-parameter exponential type potential energies \cite{Diao_et_al_2004, Olgar_2009, Lutfuoglu_Ikot_et_al_2018} and non central potentials \cite{Chen_et_al_2008, Ortakaya_2012} are examined in the KG equation.

Recently, the investigation of different physical systems in one or three dimensions have been extended to higher dimensions to describe different phenomena not only in diverse fields of physics but in quantum chemistry, too \cite{Dong_Book}. Chen \emph{et al.} examined hydrogen type atoms by employing the  Couloumb potential energy in KG equation in D-dimensions \cite{Chen_et_al_2003}. Saad \emph{et al.} applied AIM to study KG equation with unequal vector and scalar Kratzer potential energy in  D-dimensions \cite{Saad_et_al_2008}. In 2011, Hassanabadi \emph{et al.} obtained an approximate solution by employing an equal scalar and vector generalized Kratzer potential to the D-dimensioanal KG equation for any angular momentum parameter \cite{Hassanabadi_et_al_2011}. One year later, Hassanabadi \emph{et al.} examined the Eckart potential in addition to modified Hylleraas potential energy in higher dimensional relativistic equations by supersymmetric quantum mechanic  methods \cite{Hassanabadi_et_al_2012}. Ibrahim \emph{et al.} studied higher dimensional KG and Dirac equations with mixed equal scalar and vector RM potential energies by NU method \cite{Ibrahim_et_al_2012}. Ortakaya used pseudoharmonic oscillator potential energy in D-dimensional KG equation to obtain the bound state energy spectrum of $CH$, $H_2$ and $HCl$ molecules \cite{Ortakaya_2013}.  Antia \emph{et al.} defined a combined potential energy function by addition of Mobius square potential to Yukawa potential energy. Then, they employed the NU method to solve the combined potential energy in high dimensional KG equation \cite{Antia_et_al_2013}. Chen \emph{et al.} obtained the relativistic bound state energy equation by employing the improved MR potential energy in D spatial dimensions \cite{Chen_et_al_2014}. Ikot \emph{et al.} analyzed the improved MR potential energy for arbitrary angular momentum parameter in an approximate method in D-dimensions \cite{Ikot_et_al_2014}. Tan \emph{et al.} and Jia \emph{et al.} solved the D-dimensional KG equation with the improved and modified RM potential energy by employing supersymmetric WKB approximation \cite{Tan_et_al_2014, Jia_et_al_2015}. Xie \emph{et al.} examined Morse potential energy in KG equation to derive the  bound state energy equation in D spatial dimensions \cite{Xie_et_al_2015}. Ikot \emph{et al.} employed NU method to analyze an exponential type molecule potential in the KG equation in D-dimensions \cite{Ikot_Lutfuoglu_2016}. 

In last decade, the prediction of the properties of a physical system by investigating their thermodynamic functions become popular.  In this purpose, the scientist calculates the energy spectrum of the system in a relativistic or non-relativistic equation by proposing potential energy and then obtains the partition function.  Ikhdair \emph{et al.} solved the Schr\"odinger equation with the PT potential energy via AIM and discussed the thermodynamic functions \cite{Ikhdair_et_al_2013}. In 2014, Oyewumi \emph{et al.} used the shifted Deng-Fan potential energy in the non-relativistic equation to analyze the statistical properties \cite{Oyewumi_et_al_2014}. One year later, Onate  \emph{et al.} defined the combination of hyperbolical and generalized PT potential energies and solved Dirac equation. They discussed the thermodynamic properties in non relativistic limit in addition to the the spin symmetry (SS) and pseudospin symmetry (PSS) limits \cite{Onate_et_al_2015}. In 2016, Arda \emph{et al.} used the linear potential to investigate the thermodynamic quantities such as the Helmholtz free energy, and the mean energy with the specific heat function in both KG and Dirac equations \cite{Arda_et_al_2016}. Onyeaju \emph{et al.} studied the Dirac equation with the deformed Hylleraas in addition to WS potential energy and calculated the thermodynamic functions of some diatomic molecules \cite{Onyeaju_et_al_2017}. Then, Ikot \emph{et al.} discussed the thermodynamic functions of diatomic molecules by using a general molecular potential \cite{Ikot_et_al_2018}. In another paper, Valencia-Ortega and Arias-Hernandez investigated the thermodynamic properties of diatomic molecules by adopting $SO(2,1)$ anharmonic Eckart potential energy \cite{Ortega_et_al_2018}. Furthermore, Okorie with co-authors investigated thermodynamic functions by using modified Mobius square \cite{Okorie_et_al_2018_a}, modified Yukawa  \cite{Okorie_et_al_2018_b}, quadratic exponential-type \cite{Okorie_et_al_2018_c}, shifted Tietz-Wei \cite{Okorie_et_al_2018_d, Ikot_et_al_2019_e} potential energies. In 2019, one of the authors of the present paper, Ikot, with his collaborators studied the thermodynamic properties of a q-deformed quantum oscillator in the scale of minimal length \cite{Ikot_et_al_2019_ML}. The other author of the present paper, L\"utf\"uo\u{g}lu, also contributed to the field by the studies via the investigation of the generalized symmetric WS potential energy in non relativistic \cite{Lutfuoglu_et_al_2016} and relativistic equations \cite{Lutfuoglu_2019}. With the non relativistic results, they obtained the thermodynamic properties of a nucleon in relatively small \cite{Lutfuoglu_et_al_2016_ME} and big radius nuclei \cite{Lutfuoglu_et_al_2017_ME}. Then, he compared the thermodynamic functions with excluding and including the surface effects in non-relativistic \cite{Lutfuoglu_2018_1}, and relativistic regimes \cite{Lutfuoglu_2018_2}. In a very recent article, they presented the variance of the thermodynamic functions in the existence of attractive or repulsive surface interaction terms \cite{Lutfuoglu_et_al_2019}. Besides these works, thermodynamic properties of molecules and dimers are examined in several articles by taking the vibrational and rotational partition functions into account \cite{Chen_et_al_2013, Hu_et_al_2014, Jia_et_al_2017_1, Song_et_al_2017_2, Jia_et_al_2017_3, Wang_et_al_2017_4, Jia_et_al_2018_1, Jia_et_al_2018_2, Jia_et_al_2018_3, Ocak_et_al_2018, Deng_et_al_2018, Jia_et_al_2018_4, Khordad_et_al_2019, Jia_et_al_2019_ek_1, Jiang_et_al_ek_2, Jia_et_al_2019_ek_3, Jiang_et_al_ek_4}.   

Our motivation is to determine the bound state solution of the energy-dependent deformed Hulth\'en potential in D-dimensional KG equation and discuss the corresponding thermodynamic functions. Note that the energy-dependent potential energies have been investigated in both relativistic and non-relativistic wave equations since 1940 \cite{Synder_1940, Schiff_1940, Green_1962, Formanek_et_al_2004, Lombard_et_al_2007, Benchikha_et_al_2013, Benchikha_et_al_2014}. In recent times, Gupta \emph{et al.} studied the Schr\"odinger equation with energy dependent harmonic oscillator potential energy function to describe quark systems \cite{Gupta_et_al_2012}. Ikot \emph{et al.} examined energy dependent Yukawa potential energy with a Coloumb-like tensor interaction in the Dirac equation at the SS and PSS limits \cite{Ikot_Hassanabadi_2013}. Boumali \emph{et al.} examined energy dependent harmonic oscillator in Schr\"odinger \cite{Boumali_et_al_2017} and KG equation \cite{Boumali_et_al_2018} to predict the Shannon entropy and Fisher information.

The paper is organized as follows. In Sec. \ref{KGD} we define the KG equation in an arbitrary dimension with the vector and the scalar potential energy coupling. Then, we describe q-deformed energy dependent Hult\'en potential energy and obtain the radial wave function solution by employing a Greene-Aldrich approach to the centrifugal term. Furthermore, we derive the quantization condition. Before we end the section, we briefly give the normalization method in an energy dependent potential energy case. In Sec. \ref{Thermo} we state the thermodynamic functions such as Helmholtz free energy, entropy, internal energy, and specific heat. Then, in Sec. \ref{RD} we use the Newton-Raphson method to calculate energy spectra for various dimensions in the EMES, EMOS, pure vector and scalar limits. Moreover, we obtain the thermodynamic functions from the partition function. We demonstrate those functions within a comparison. In Sec. \ref{Concl} we conclude the paper.

\section{Solutions of the Klein-Gordon equation in D-Dimensions} \label{KGD}

We start by expressing the KG equation in $D$ spatial dimensions with
\begin{eqnarray}
\big[\hat{p}^\mu\hat{p}_\mu-(m_0c)^2\big]\phi(\vec{r},t)=0. \hspace{20mm} \mu=0,1,\cdots, D.
\end{eqnarray}
Here,  we use $\hat{p}^\mu$, $c$  and $m_0$  to denote the $(D+1)$ momentum vector, the speed of light and the rest mass of the particle, respectively. Then, we employ a minimal coupling of the momentum vector to a $(D+1)$ vector potential. Among the components of the vector potential, we only assume that the time component have non-zero value. This component, $V_v$, is called as "the vector potential" in the literature. In addition, we use a scalar potential, $V_s$, coupling to the rest mass parameter term.

In this manuscript, we investigate the solution of the spherical symmetric potential energies that are time-independent. Therefore, we can separate the wave function into time and spatial components. Then, we decompose the spatial part of the wave function into radial and angular parts by employing the spherical symmetric nature of the potential energies. Finally, we obtain the radial equation as follows. 
\begin{eqnarray}
  \Bigg[\frac{d^2 }{d r^2}-\frac{\gamma}{r^2}+\frac{1}{\hbar^2c^2}\Big[ \big(E -V_v \big)^2- \big(m_0c^2 + g V_s\big)^2\Big]\Bigg]\chi(r) &=& 0, \label{1} \,\,\,\,\,\,\,\,\,\,\,\,\,\,\,\,\,\,\,\,
\end{eqnarray}
Here $\gamma\equiv \frac{(D+2l-1)(D+2l-3)}{4}$, and $l$ denotes the angular momentum quantum number. Furthermore, $\hbar$ represents the Planck constant, and $g$ is the coupling constant that is nearly equal to one in the strong regime.  Note that $\chi(r)\equiv rR(r)$. In the rest of the article, we will use the  natural units where $\hbar=c=1$.

\subsection{Bound State Solutions}

We examine q-deformed energy dependent  vector and scalar Hulth\'{e}n potential energy wells 
\begin{eqnarray}
V_v(r)&=&-V_0\frac{(1+aE)e^{-\delta r}}{1-q e^{-\delta r}}, \label{2}\\
g V_s(r)&=&-S_0\frac{(1+aE)e^{-\delta r}}{1-q e^{-\delta r}}. \label{3}
\end{eqnarray}
where $V_0$, $S_0$, $a$   and  $\delta$ are the vector potential depth, scalar potential depth, energy slope parameter and the screening parameters, respectively.

In order to deal with the centrifugal term we adopt the Greene-Aldrich approximation scheme \cite{Greene_et_al_1976}
\begin{eqnarray}
\frac{1}{r^2} \approx \delta^2 \frac{e^{-2 \delta r}}{\Big(1-q e^{-\delta r}\Big)^2}. \label{4}
\end{eqnarray}
Here,  $\delta r<1$ and $q\simeq 1$. Note that for the validity, the deformation parameter value should not be higher than $1$. Then, we substitute Eq.~(\ref{2}), Eq.~(\ref{3}) and Eq.~(\ref{4}) into Eq.~(\ref{1}) and we get
\begin{eqnarray}
&&\Bigg\{\frac{d^2}{dr^2}+ E^2-m^2+2(E V_0+mS_0)\frac{(1+aE)e^{-\delta r}}{1-q e^{-\delta r}}\nonumber \\&+&(V_0^2-S_0^2)\frac{(1+aE)^2e^{-2\delta r}}{(1-q e^{-\delta r})^2}- \gamma\delta^2 \frac{e^{-2 \delta r}}{(1-q e^{-\delta r})^2}\Bigg\}\chi(r)=0. \label{KG2}
\end{eqnarray}
We introduce a new coordinate transformation of the form
$z \equiv \big(1-qe^{-\delta r}\big)^{-1} $, 
and adopt the following abbreviations
\begin{eqnarray}
\varepsilon&\equiv&(1+aE), \\
-\alpha^2&\equiv&\frac{E^2-m^2}{\delta^2}, \\
\sigma^2&\equiv&\frac{2\varepsilon(EV_0+mS_0)}{q\delta^2},\\
\beta^2&\equiv&\frac{1}{q^2} \bigg[\frac{(V_0^2-S_0^2)\varepsilon^2}{\delta^2}-\gamma \bigg].
\end{eqnarray}
We get
\begin{eqnarray}
\Bigg\{z(z-1)\frac{d^2}{dz^2}+(2z-1)\frac{d}{dz}+\beta^2- \frac{\alpha^2}{z-1}+\frac{\alpha^2+\sigma^2-\beta^2}{z}\Bigg\}\phi(z)=0. \label{KG3}
\end{eqnarray}
Then, we propose the following ansatz  
\begin{eqnarray}
\phi(z)&\equiv& z^\mu (z-1)^\nu u(z).
\end{eqnarray}
where
\begin{eqnarray}
\mu^2&=&\alpha^2+\sigma^2-\beta^2,\label{mu}\\
\nu^2&=&\alpha^2,\label{nu}\\
\theta^2&=&\frac{1}{4}-\beta^2.
\end{eqnarray}
We find that Eq.~(\ref{KG3}) turns into the following form
\begin{eqnarray}
&&z(1-z)\frac{d^2u(z)}{dz^2}+\bigg[(1+2\mu)-z(2\mu+2\nu+2)\bigg]\frac{du(z)}{dz}
\nonumber \\
&&-\bigg[(\mu+\nu+\frac{1}{2}+\theta)(\mu+\nu+\frac{1}{2}-\theta)\bigg]u(z)=0.
\end{eqnarray}
The solution can be expressed in terms of the hypergeometric functions $_2F_1$
\begin{eqnarray}
\phi(z)&=&N_1 z^\mu (z-1)^\nu\,  
_2F_1(d,b,c,z)\nonumber \\
&&+ N_2 z^{-\mu} (z-1)^\nu \,_2F_1(1+d-c,1+b-c,2-c,z),
\end{eqnarray}
where
\begin{eqnarray}
d&\equiv& \mu+\nu+\frac{1}{2}-\theta, \\
b&\equiv&\mu+\nu+\frac{1}{2}+\theta, \\
c&\equiv&1+2\mu.
\end{eqnarray}

\subsection{Quantization} 
In this subsection, we take into account the boundary condition that dictates the radial wave function should go to zero at infinity. In that limit, the transformed coordinate $z$ goes to $1$. Therefore, we need to determine the behaviour of the hypergeometric function initially. We employ  the following well-known property of the hypergeometric function \cite{Abramowitz_et_al_Book}
\begin{eqnarray}
&&_2F_1(d,b,c,t)= \frac{\Gamma(c)\Gamma(c-d-b)}{\Gamma(c-d)\Gamma(c-b)} {}_2F_1(d,b,d+b-c+1,1-t)\nonumber\\ &&+(1-t)^{c-d-b} \frac{\Gamma(c)\Gamma(d+b-c)}{\Gamma(d)\Gamma(b)}{}_2F_1(c-d,c-b,c-d-b+1,1-t).\,\,
\end{eqnarray}
Then, we find  
\begin{eqnarray}
&&\phi(z)=N_1 z^\mu (z-1)^\nu\Bigg[\frac{\Gamma(c)\Gamma(c-d-b)}{\Gamma(c-d)\Gamma(c-b)} {}_2F_1(d,b,d+b-c+1,1-z)\Bigg] \nonumber\\ 
&&+N_1e^{-2\pi i \nu}z^\mu (z-1)^{-\nu}\Bigg[ \frac{\Gamma(c)\Gamma(d+b-c)}{\Gamma(d)\Gamma(b)}{}_2F_1(c-d,c-b,c-d-b+1,1-z)\Bigg] 
\nonumber \\
&&+ N_2 z^{-\mu} (z-1)^\nu 
\Bigg[\frac{\Gamma(2-c)\Gamma(c-d-b)}{\Gamma(1-d)\Gamma(1-b)}{}_2F_1(d-c+1,b-c+1,d+b-c,1-z)\Bigg]\,\,\,\,\,\,\nonumber\\
&&+ N_2e^{-2\pi i \nu} z^{-\mu} (z-1)^{-\nu} \Bigg[\frac{\Gamma(2-c)\Gamma(d+b-c)}{\Gamma(d-c+1)\Gamma(b-c+1)}{}_2F_1(1-d,1-b,c-d-b,1-z)\Bigg].\,\,\,\,\,\,\,\,\,\,\,\,\,\,\,\,\,\,\
\end{eqnarray}
After this identical transformation of the hypergeometric functions, the result values are equal to 1. Consequently, we get
\begin{eqnarray}
 \Big[N_1 \Gamma_1+N_2\Gamma_3\Big]   (z-1)^\nu  +\Big[N_1\Gamma_2+N_2 \Gamma_4 \Big] e^{-2\pi i \nu}  (z-1)^{-\nu}=0,
\end{eqnarray}
where 
\begin{eqnarray}
\Gamma_1&\equiv& \frac{\Gamma(1+2\mu)\Gamma(-2\nu)}{\Gamma(\mu-\nu+\frac{1}{2}+\theta)\Gamma(\mu-\nu+\frac{1}{2}-\theta)},  \\
\Gamma_2&\equiv&  \frac{\Gamma(1+2\mu)\Gamma(2\nu)}{\Gamma(\mu+\nu+\frac{1}{2}-\theta)\Gamma(\mu+\nu+\frac{1}{2}+\theta)},  \\
\Gamma_3&\equiv& \frac{\Gamma (1-2\mu)\Gamma(-2\nu)}{\Gamma(-\mu-\nu+\frac{1}{2}+\theta)\Gamma(-\mu-\nu+\frac{1}{2}-\theta)},  \\
\Gamma_4&\equiv& \frac{\Gamma(1-2\mu)\Gamma(2\nu)}{\Gamma(-\mu+\nu+\frac{1}{2}-\theta)\Gamma(-\mu+\nu+\frac{1}{2}+\theta)}.
\end{eqnarray}
We assume that $\nu=\alpha$. We are obliged to take $N_2=0$ and $\Gamma_2=0$ to avoid the singularity. We use the definition of the reciprocal of gamma function for negative integer as given  in \cite{Abramowitz_et_al_Book},
\begin{eqnarray}
\lim_{z\rightarrow n} \frac{1}{\Gamma(-z)}=0, \,\,\,\,\,\,\,\,\,\,\,\, n=0,1,2,\cdots
\end{eqnarray}
to eradicate $\Gamma_2$. 
Although either $\mu+\nu+\frac{1}{2}-\theta$ or 
$\mu+\nu+\frac{1}{2}+\theta$ can be chosen to be equal to $-n$, the symmetric structure of the wave functions under exchange of both parameters lead to obtain the same solution. We use the condition
\begin{eqnarray}
\mu+\nu+\frac{1}{2}-\theta&=&-n, \label{firstround}
\end{eqnarray}
and we obtain 
\begin{eqnarray}
\sqrt{m^2-E^2}&=&\frac{\delta}{2}\Bigg[\Sigma-n+\frac{\beta^2-\sigma^2}{\Sigma-n}\Bigg].
\end{eqnarray}
where  
\begin{eqnarray}
\Sigma&=& \sqrt{\frac{1}{4}- \frac{(V_0^2-S_0^2)\varepsilon^2}{q^2\delta^2}+\frac{\gamma}{q^2}}-\frac{1}{2},
\end{eqnarray}
We find the  unnormalized radial wave function as follows
\begin{eqnarray}
R_n(r)&=& \frac{N_1}{r} \Big(\frac{1}{1-qe^{-\delta r}}\Big)^\mu \Big(\frac{q}{e^{\delta r}-q}\Big)^\nu \nonumber \\
&\times& _2F_1\Big(-n,n+2\mu+2\nu+1,1+2\mu,\frac{1}{1-qe^{-\delta r}} \Big). 
\end{eqnarray}

\subsection{The normalization of the radial wave function with energy dependent potential energies}
A. Benchikha \emph{et al.} examined the energy dependent potential energy in non-relativistic \cite{Benchikha_et_al_2013}  and  relativistic \cite{Benchikha_et_al_2014}  equations. They modified the well-known probability density definition for the KG equation with the following expression  
\begin{eqnarray}
\Big|\chi_E(x)\Big|^2\Bigg(1-\frac{\partial^2 }{\partial E^2}\Big[E^2-\big(E-V(x,E)\big)^2+\big(M+S(x,E)\big)^2\Big]\Bigg)\Bigg|_{E=E_n} .
\end{eqnarray}
Consequently, in the problem one can calculate the normalization constant as follows
\begin{eqnarray}
\frac{1}{|N|^2}&=&\int_0^\infty r^2 dr  \Bigg(1+\frac{4aV_0e^{-\delta r}}{1-q e^{-\delta r}}+\frac{2a^2(V_0^2-S_0^2)e^{-2\delta r}}{(1-q e^{-\delta r})^2}\Bigg)|R(r)|^2.
\end{eqnarray}
Here, we skip calculating the normalization constant since it does not exist in our main motivation.

\section{Thermodynamic functions} \label{Thermo}
One way to examine the thermodynamic properties of a physical system is to use the partition function.  In the canonical ensemble, for a system that is in an equilibrium state, the partition function is defined with 
\begin{eqnarray}
Z(\beta_T)&=&\sum_{n=0} e^{-\beta_T E_n}. \label{PF}
\end{eqnarray}
Here, $E_n$ represents the available microstate energy values.  $\beta_T$ is the reciprocal temperature function and it is inversely proportional to the multiplication of the Boltzmann constant with the absolute temperature. Thermodynamic functions such as  Helmholtz free energy, $F(\beta_T)$, entropy, $S(\beta_T)$, internal energy, $U(\beta_T)$,  and specific heat, $C_v(\beta_T)$, functions are obtained from the partition function as follows
\begin{eqnarray}
F(\beta_T)&\equiv&-\frac{1}{\beta_T} \ln Z(\beta_T), \label{HFE}\\
S(\beta_T)&=&-k_B\frac{\partial F(\beta_T)}{\partial \beta_T}, \label{SE} \\
U(\beta_T)&\equiv&-\frac{\partial}{\partial \beta_T} \ln Z(\beta_T), \label{IE}\\
C_v(\beta_T)&=& k_B\frac{\partial U(\beta_T)}{\partial \beta_T}. \label{SH}
\end{eqnarray} 

\section{Results and discussions} \label{RD}
In this section, we construct the thermodynamic functions just after we calculate the energy spectra in different limits and dimensions. To calculate the energy spectra we solve the quantization condition numerically by the use of the Newton-Raphson method in  the EMES limit, $V_0=S_0$, in the EMOS limit,  $V_0=-S_0$,  in the pure vector limit, $S_0=0$, and in the pure scalar limit, $V_0=0$. Note that, since we study with the natural units, all units of the parameters of the system can be expressed in terms of energy or reciprocal energy. There are some parameters that are always kept as a constant in all limits, for instance, the mass and the deformation parameter. Both of them are equal to one. There are some other parameters,  which we assign different values, i.e., $a$ parameter, which is the measure of the energy dependence of the potential energy, is assumed to be equal to $1$ $(1/E)$, $0$, and $-1$ $(1/E)$.  Note that, we calculate the spectra only in $3$, $4$, and $5$ dimensions. 

In the second part of this section, namely in subsection \ref{thermoapp}, we employ the obtained energy spectra to discuss the thermodynamic functions of the system.

\subsection{EMES limit}
We assume the energy depth parameters have equal values as given, $V_0=S_0=2$ $(E)$. Moreover, the slope parameter is equal to $0.01$ $(E)$.  We tabulate the energy spectra  in three dimensions in Table \ref{TabSS3}, in four dimensions in Table \ref{TabSS4}, and in five dimensions in Table \ref{TabSS5}, respectively.

\subsection{EMOS limit}
In this limit, the energy depth parameters have negatively equal values. Here, we assume  $V_0=2$ $(E)$ and $S_0=-2$ $(E)$. Alike EMES limit, we choose the slope parameter to be  equal to $0.01$ $(E)$.  Then, we present the energy spectra  in three dimensions in Table \ref{TabPSS3}, in four dimensions in Table \ref{TabPSS4}, and in five dimensions in Table \ref{TabPSS5}, respectively.

We see that when the energy dependence is fixed with $\varepsilon=1-E$, ($a=-1$ $(1/E)$), most of the eigenvalues in the energy spectrum cannot be calculated. Therefore, we decide to calculate the spectrum for higher values of $l$ parameter. Surprisingly, unlike  $\varepsilon=1+E$ case, the values of $l$ parameter are not limited. In three, four and five dimensions we repeat the calculations and present them in Table \ref{TabPSS3n}, in Table \ref{TabPSS4n}, and in Table \ref{TabPSS5n}, respectively.  We conclude that as the values of parameter $l$ increase, energy eigenvalues converge.

\subsection{Pure vector limit}
In this limit, the scalar potential energy term is taken to be zero. Alike the previous limits, we assume that the $V_0=2$ $(E)$. Unlike, we examine two different values of the slope parameter and tabulate it in Table \ref{TabPV}. We find that when the energy dependence is lost, only one value of energy appears in the spectrum. 

\subsection{Pure scalar limit}
In this limit, the scalar potential energy term is equal to  $2(E)$, while the vector potential energy term is zero.  Alike the pure vector limit, we examine two different values of the slope parameter. We present the results in Table \ref{TabPS}. We find that there is only one energy eigenvalue in pure scalar spectra unlike the vector limit. Moreover, when the potential energy does not depend on energy, ground state energy eigenvalues do not occur. 

\subsection{Thermodynamic properties}  \label{thermoapp}
In this subsection, we use the EMES limit case results to examine the thermodynamic properties of the system.  Therefore, we only employ  Table \ref{TabSS3}, Table \ref{TabSS4}, and Table \ref{TabSS5} to construct the partition function. 

\begin{figure}[!htb]
\centering
\includegraphics[totalheight=0.45\textheight,clip=true]{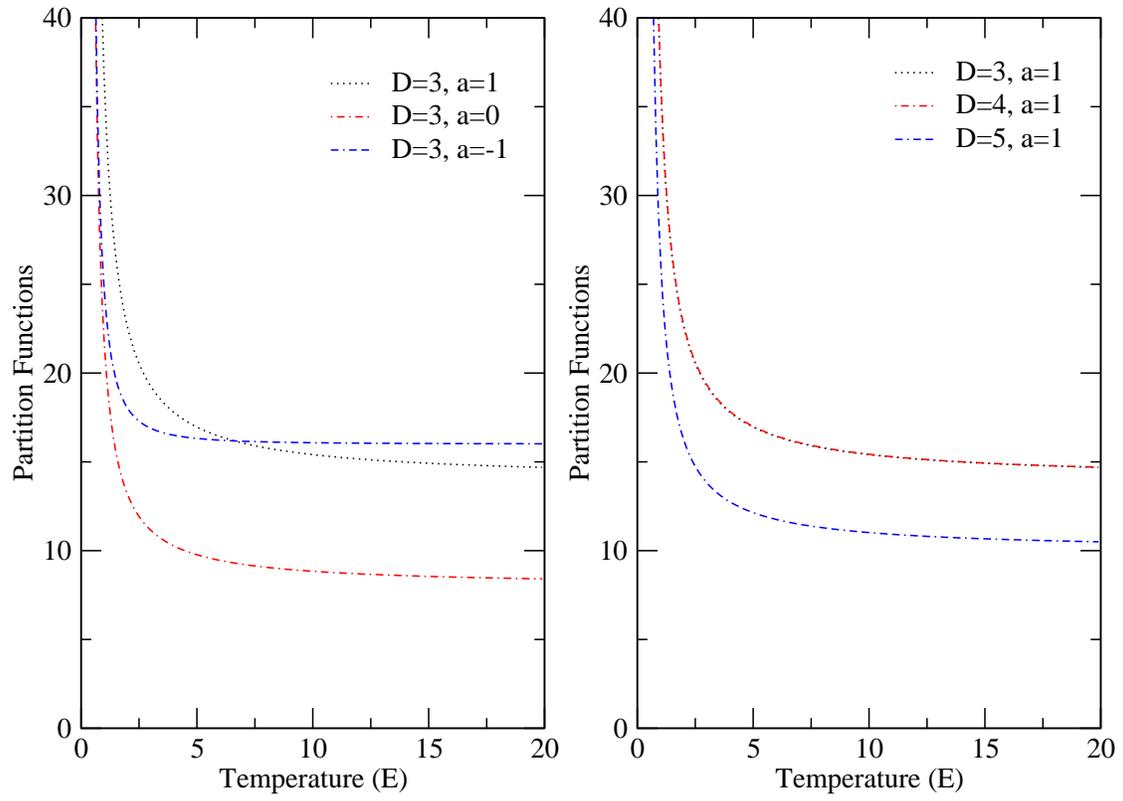}
   \caption{Comparison of the partition functions versus the temperature in the EMES limit.} \label{fig1:PF}
\end{figure}

First, we use of the energy eigenvalues  for $a=1$ $(1/E)$, $a=0$ $(1/E)$, and $a=-1$ $(1/E)$  in three dimensions from Table \ref{TabSS3}. We calculate the partition functions from Eq. (\ref{PF}) and plot them in the first column of Fig. \ref{fig1:PF}. Then, we use the energy spectra in three, four and five dimensions for the $a=1$ $(1/E)$ case from Table \ref{TabSS3}, Table \ref{TabSS4}, and Table \ref{TabSS5}. We present the plot of the partition functions in the second column of Fig. \ref{fig1:PF}. We see that the partition functions in three and four dimensions overlap.

\begin{figure}[!htb]
\centering
\includegraphics[totalheight=0.45\textheight,clip=true]{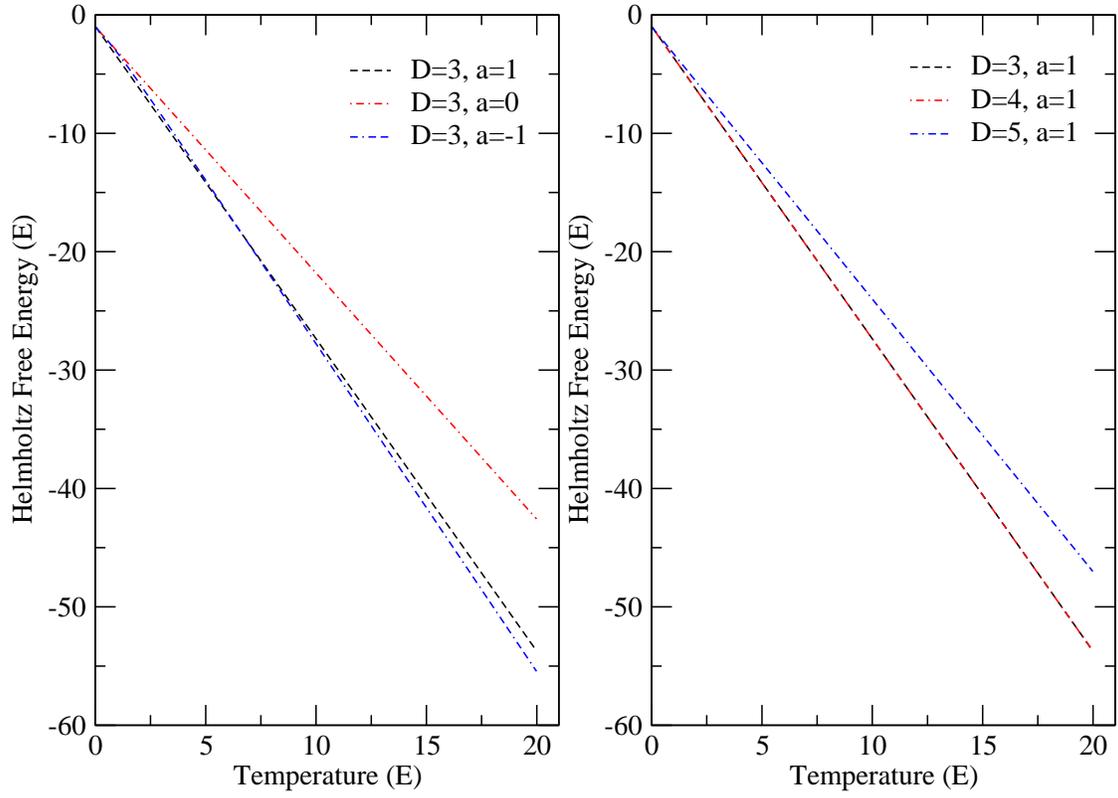}
   \caption{Comparison of the Helmholtz free energy functions versus the temperature in the EMES limit.} \label{fig2:HFE}
\end{figure}

We obtain the Helmholtz free energy functions by employing Eq. (\ref{HFE}). We demonstrate the three-dimensional results in the first column of Fig. \ref{fig2:HFE}. We see that Helmholtz free energy function for energy-dependent function cases have a very close appearance.  We put forth the higher dimensional cases results in the second column of Fig. \ref{fig2:HFE}. We find out that the overlapping of the thermodynamic functions is still valid.  

\begin{figure}[!htb]
\centering
\includegraphics[totalheight=0.45\textheight,clip=true]{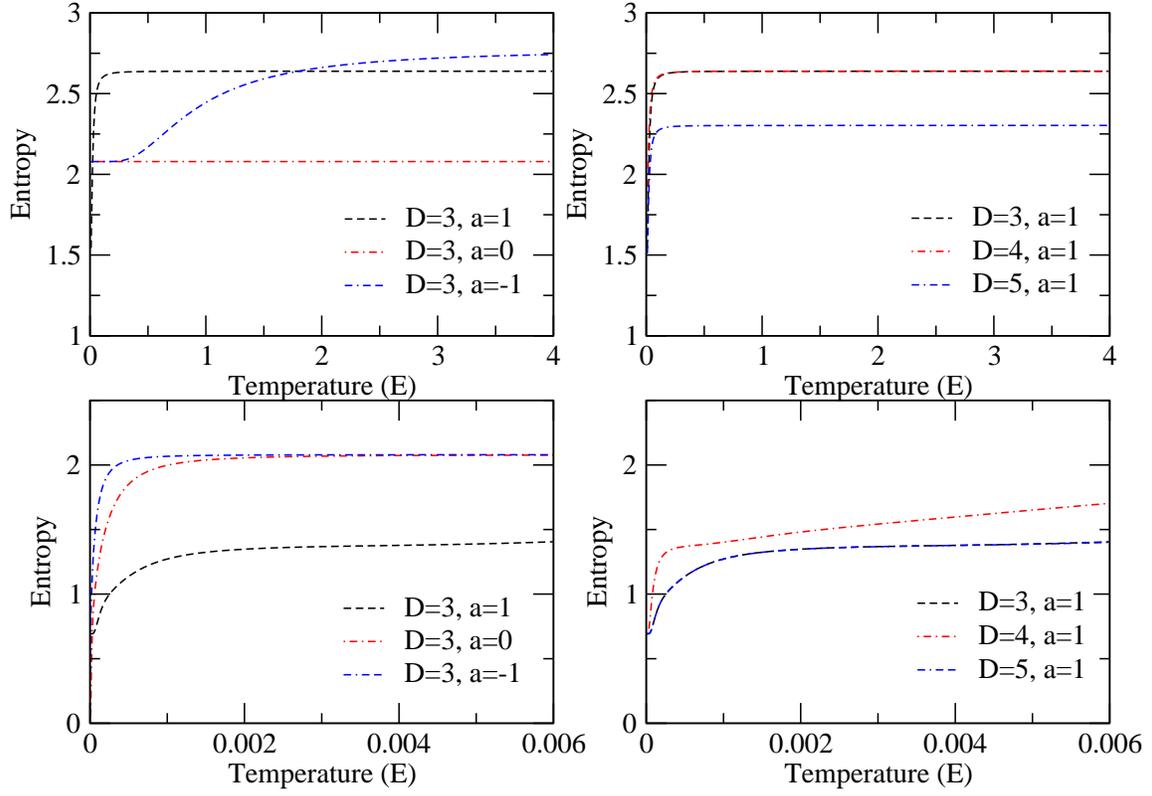}
   \caption{Comparison of the entropy functions versus the temperature in the EMES limit.} \label{fig3:EF}
\end{figure}

We derive the entropy function from the Helmholtz free energy via  Eq. (\ref{SE}). We show entropy functions versus lower temperature and relatively higher temperature in Fig. \ref{fig3:EF}. The entropy function in three dimensions behaves like the entropy function of five dimensions at low temperatures, while it behaves like the entropy function obtained in four dimensions at relatively high temperatures. Another finding is, in three dimensions at a lower temperature the entropy functions for $a=0$ and $a=-1$ $(1/E)$ case act similar to each other while at a relatively high temperature not.

\begin{figure}[!htb]
\centering
\includegraphics[totalheight=0.45\textheight,clip=true]{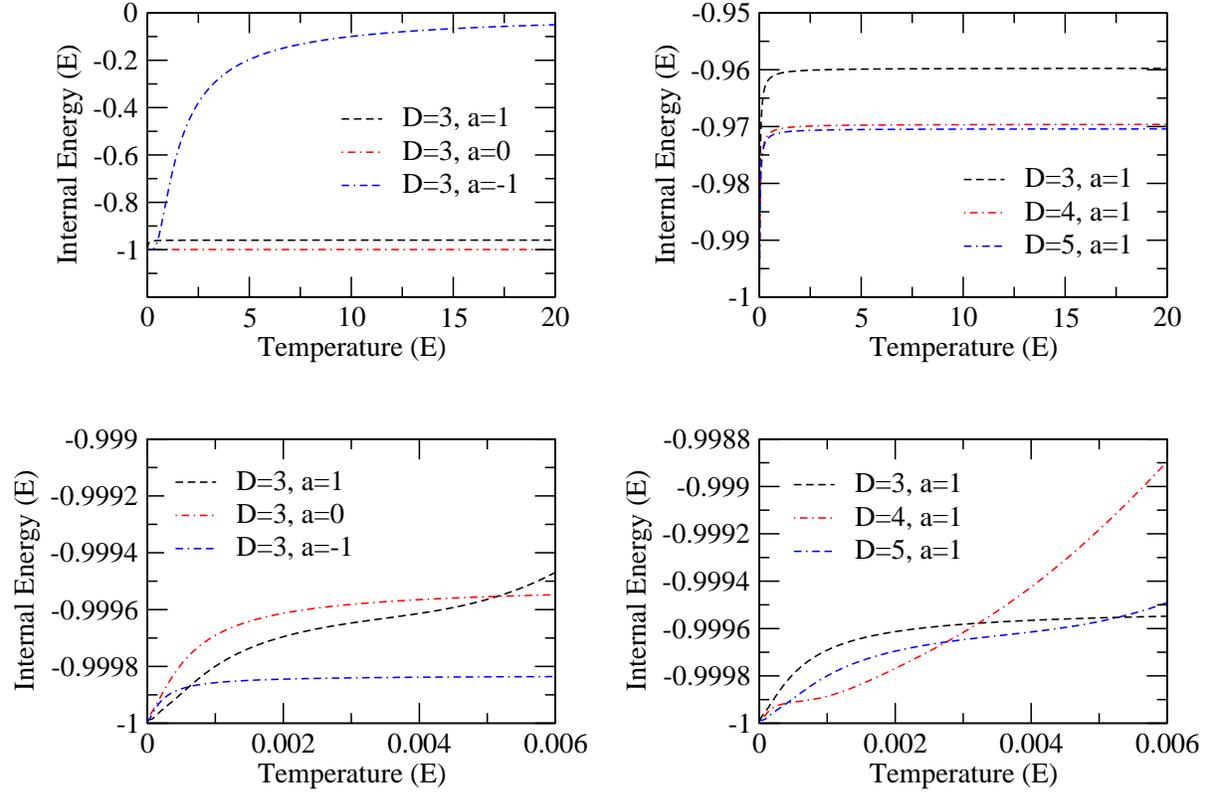}
   \caption{Comparison of the internal energy functions versus the temperature in the EMES limit.} \label{fig4:IEF}
\end{figure}

Then, we use Eq. (\ref{IE}) to compute the internal energy functions. We present internal energy functions in Fig. \ref{fig4:IEF} versus temperature. We conclude that mean energy values are compatible with the results.

\begin{figure}[!htb]
\centering
\includegraphics[totalheight=0.45\textheight,clip=true]{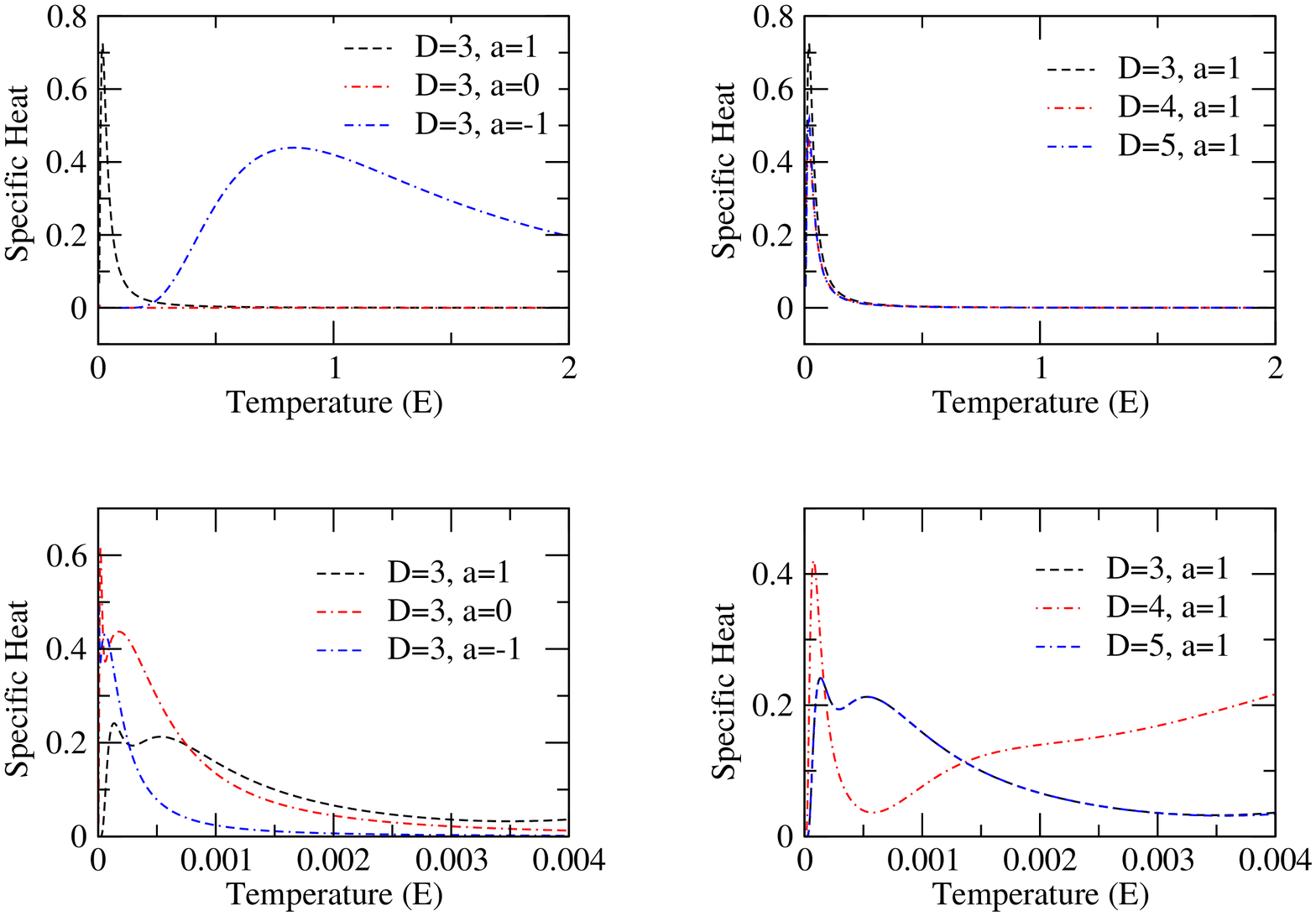}
   \caption{Comparison of the specific heat functions versus the temperature in the EMES limit.} \label{fig5:SHF}
\end{figure}

Finally, we achieve the specific heat function with the help of Eq. (\ref{SH}). We present them in Fig. \ref{fig5:SHF} versus temperature. We conclude that at a relatively higher temperature in all dimensions the characteristic of the functions for $a=1$ $(1/E)$ case, remains the same. On the other hand, in three dimensions, the specific heat function of $a=-1$ $(1/E)$ case,  differs from others.

\section{Conclusion} \label{Concl}
In this article, we investigated the bound state solutions of a mixed vector and scalar energy-dependent deformed Hulth\'en potential in the KG equation in arbitrary dimension. We obtained a transcendental equation which yields to the quantization of the energy eigenvalues by the use of the necessary boundary conditions. Then, we employed the Newton-Raphson method to calculate energy spectra in the limits of the EMES, EMOS, pure vector and pure scalar. Finally, we used the canonical partition function definition and derived other thermodynamic functions, such as Helmholtz free energy, entropy, internal energy, and specific heat. Then, we discussed thermodynamic properties with energy dependency and dimensional effects.

\section*{Acknowledgment}
The authors thank the kind reviewers of the article for the positive comments and suggestions that leads to an improvement in the quality of the article.  B.C. L\"utf\"uo\u{g}lu is supported by the Turkish Science and Research Council (TUBITAK), and Akdeniz University.

\newpage
\begin{table}[ht]
    \centering
    \begin{tabular}{|p{5mm}|p{5mm}||p{20mm}|p{20mm}||p{20mm}||p{20mm}|p{20mm}|| }
\hline
\multicolumn{7}{|c|}{$E_{nl}$ $(E)$} \\
 \hline
  $n$   & $l$&    \multicolumn{2}{|c||}{$a=1$ $(1/E)$}&$a=0$ $(1/E)$&\multicolumn{2}{|c||}{$a=-1$ $(1/E)$} \\
 \hline
$1$    & $0$    &$-0.962938$&  $none     $     & $-0.999907$    &$-0.999967$    &$0.999967$\\
\hline
$2$    & $0$    &$-0.940966$&  $none     $     & $-0.999627$    &$-0.999869$    &$0.999869$\\
       & $1$    &$-0.963843$&  $-0.999987$     & $-0.999988$    &$-0.999997$    &$0.999997$\\
\hline
$3$    & $0$    &$-0.922427$   &$none     $    & $-0.999160$    &$-0.999705$    &$0.999705$\\
       & $1$    &$-0.941524$   &$none     $    & $-0.999709$    &$-0.999907$    &$0.999907$\\
       & $2$    &$-0.965764$   &$-0.999687$    & $none     $    &$none     $    &$none    $\\
\hline
$4$   & $0$    &$-0.905797$    &$none     $    & $-0.998508$    &$-0.999476$    &$0.999476$\\
      & $1$    &$-0.922848$    &$none     $    & $-0.999240$    &$-0.999742$    &$0.999742$\\
      & $2$    &$-0.942664$    &$-0.999987$    & $-0.999950$    &$-0.999987$    &$0.999987$\\
      & $3$    &$-0.969016$    &$-0.998460$    & $none     $    &$none     $    &$none    $\\
\hline
\end{tabular} 
    \caption{Energy spectrum for the EMES limit in three dimensions.}
    \label{TabSS3}
\end{table}

\newpage
\begin{table}[ht]
    \centering
\begin{tabular}{|p{5mm}|p{5mm}||p{20mm}|p{20mm}||p{20mm}||p{20mm}|p{20mm}|| }
\hline
\multicolumn{7}{|c|}{$E_{nl}$ $(E)$} \\
 \hline
$n$   & $l$    &\multicolumn{2}{|c||}{$a=1$ $(1/E)$}&$a=0$ $(1/E)$&\multicolumn{2}{|c||}{$a=-1$ $(1/E)$} \\
 \hline
$1$   & $0$    &$-0.977245$   &$-0.999987$  &$-0.999997$    &$none$       &$none$      \\
\hline
$2$   & $0$    &$-0.951600$   &$none$       &$-0.999820$    &$-0.999940$  &$0.999940$  \\
      & $1$    &$-0.979644$   &$-0.999382$  &$none     $    &$none    $   &$none    $  \\
\hline
$3$   & $0$    &$-0.931576$   &$none$       &$-0.999447$    &$-0.999809$  &$0.999809$  \\
      & $1$    &$-0.952636$   &$-0.999987$  &$-0.999972$    &$-0.999993$  &$0.999993$  \\
      & $2$    &$-0.985401$   &$-0.995733$  &$none$         &$none$       &$none$      \\
\hline
$4$   & $0$    &$-0.914064$   &$none     $  &$-0.998887$   &$-0.999613$    &$0.999613$  \\
      & $1$    &$-0.932297$   &$none$       &$-0.999572$   &$-0.999867$    &$0.999867$  \\
      & $2$    &$-0.954443$   &$-0.999765$  &$none     $   &$none     $    &$none    $  \\
      & $3$    &$none     $   &$none     $  &$none     $   &$none     $    &$none    $   \\
\hline
\end{tabular}
\caption{Energy spectrum for the EMES limit in four dimensions.}
\label{TabSS4}
\end{table}

\newpage
\begin{table}[ht]
    \centering
\begin{tabular}{|p{5mm}|p{5mm}||p{20mm}|p{20mm}||p{20mm}||p{20mm}|p{20mm}|| }
\hline
\multicolumn{7}{|c|}{$E_{nl}$ $(E)$} \\
 \hline
$n$   & $l$    &\multicolumn{2}{|c||}{$a=1$ $(1/E)$}&$a=0$ $(1/E)$&\multicolumn{2}{|c||}{$a=-1$ $(1/E)$} \\
 \hline
$1$   & $0$    &$none     $   &$none     $  &$none     $    &$none     $   &$none    $  \\
\hline
$2$   & $0$    &$-0.963843$   &$-0.999987$  &$-0.999988$    &$-0.999997$   &$0.999997$  \\
      & $1$    &$none     $   &$none     $  &$none     $    &$none     $   &$none    $  \\
\hline
$3$   & $0$    &$-0.941524$   &$none$       &$-0.999709$    &$-0.999907$   &$0.999907$  \\
      & $1$    &$-0.965764$   &$-0.999687$  &$none     $    &$none     $   &$none    $  \\
      & $2$    &$none     $   &$none     $  &$none     $   &$none      $   &$none    $   \\
\hline
$4$   & $0$    &$-0.922848$   &$none     $  &$-0.999240$   &$-0.999742$    &$0.999742$  \\
      & $1$    &$-0.942664$   &$-0.999987$  &$-0.999950$   &$-0.999987$    &$0.999987$  \\
      & $2$    &$-0.969016$   &$-0.998460$  &$none     $   &$none     $    &$none    $  \\
      & $3$    &$none     $   &$none     $  &$none     $   &$none     $    &$none    $   \\
\hline
\end{tabular}
\caption{Energy spectrum for the EMES limit in five dimensions.}
\label{TabSS5}
\end{table}

\newpage
\begin{table}[ht]
    \centering
    \begin{tabular}{|p{5mm}|p{5mm}||p{20mm}|p{20mm}||p{20mm}||p{25mm}|| }
\hline
\multicolumn{6}{|c|}{$E_{nl}$ $(E)$} \\
 \hline
$n$    & $l$    & \multicolumn{2}{|c||}{$a=1$ $(1/E)$} &$a=0$ $(1/E)$ &$a=-1$ $(1/E)$ \\
 \hline
$1$    & $0$    &$none     $   &$none     $    & $none     $    &$none     $    \\
\hline
$2$    & $0$    &$none     $   &$none     $    & $none     $    &$none     $    \\
       & $1$    &$-0.999995$   &$0.999995 $    & $0.999993 $    &$0.999688 $    \\
\hline
$3$    & $0$    &$none     $   &$none     $    & $none     $    &$none     $    \\
       & $1$    &$none     $   &$none     $    & $none     $    &$none     $    \\
       & $2$    &$-0.999960$   &$0.999960 $    & $0.999933 $    &$0.999688 $    \\
\hline
$4$   & $0$    &$none     $    &$none     $    & $none     $    &$none     $    \\
      & $1$    &$none     $    &$none     $    & $none     $    &$none     $    \\
      & $2$    &$-0.999993$    &$0.999993 $    & $0.999991 $    &$0.999988 $    \\
      & $3$    &$-0.999898$    &$0.999898 $    & $0.999820 $    &$0.998511 $    \\
\hline
\end{tabular} 
    \caption{Energy spectrum for the EMOS limit in three dimensions.}
    \label{TabPSS3}
\end{table}

\newpage
\begin{table}[ht]
    \centering
    \begin{tabular}{|p{5mm}|p{5mm}||p{20mm}|p{20mm}||p{20mm}||p{25mm}|| }
\hline
\multicolumn{6}{|c|}{$E_{nl}$ $(E)$} \\
 \hline
$n$    & $l$    & \multicolumn{2}{|c||}{$a=1$ $(1/E)$} &$a=0$ $(1/E)$&$a=-1$ $(1/E)$ \\
 \hline
$1$    & $0$    &$-0,999997$   &$0,999997 $    & $0,999995 $    &$0,999988 $    \\
\hline
$2$    & $0$    &$none     $   &$none     $    & $none     $    &$none     $    \\
       & $1$    &$-0,999966$   &$0,999966 $    & $0,999940 $    &$0,999392 $    \\
\hline
$3$    & $0$    &$none     $   &$none     $    & $none     $    &$none     $    \\
       & $1$    &$-0.999994$   &$0.999994 $    & $0.999992 $    &$0.999988 $    \\
       & $2$    &$-0.999910$   &$0.999910 $    & $0.999833 $    &$0.996735 $    \\
\hline
$4$   & $0$    &$none     $    &$none     $    & $none     $    &$none     $    \\
      & $1$    &$none     $    &$none     $    & $none     $    &$none     $    \\
      & $2$    &$-0.999955$    &$0.999955 $    & $0.999928 $    &$0.999765 $    \\
      & $3$    &$-0.999829$    &$0.999829 $    & $0.999676 $    &$0.991779 $    \\
\hline
\end{tabular} 
\caption{Energy spectrum for the EMOS limit in four dimensions.}
\label{TabPSS4}
\end{table}

\newpage
\begin{table}[ht]
    \centering
    \begin{tabular}{|p{5mm}|p{5mm}||p{20mm}|p{20mm}||p{20mm}||p{25mm}|| }
\hline
\multicolumn{6}{|c|}{$E_{nl}$ $(E)$} \\
 \hline
$n$    & $l$    & \multicolumn{2}{|c||}{$a=1$ $(1/E)$} &$a=0$ $(1/E)$&$a=-1$ $(1/E)$ \\
 \hline
$1$    & $0$    &$none     $   &$none     $    & $none     $    &$none     $    \\
\hline
$2$    & $0$    &$-0.999995$   &$0.999995 $    & $0.999993 $    &$0.999988 $    \\
       & $1$    &$none     $   &$none     $    & $none     $    &$none     $    \\
\hline
$3$    & $0$    &$none     $   &$none     $    & $none     $    &$none     $    \\
       & $1$    &$-0.999960$   &$0.999960 $    & $0.999933 $    &$0.999688 $    \\
       & $2$    &$none     $   &$none     $    & $none     $    &$none     $    \\
\hline
$4$   & $0$    &$none     $    &$none     $    & $none     $    &$none     $    \\
      & $1$    &$-0.999993$    &$0.999993 $    & $0.999991 $    &$0.999987 $    \\
      & $2$    &$-0.999898$    &$0.999898 $    & $0.999820 $    &$0.998511 $    \\
      & $3$    &$none     $    &$none     $    & $none     $    &$none     $    \\
\hline
\end{tabular} 
\caption{Energy spectrum for the EMOS limit in five dimensions.}
\label{TabPSS5}
\end{table}

\newpage
\begin{table}[ht]
    \centering
    \begin{tabular}{|p{10mm}||p{25mm}||p{25mm}||p{25mm}||p{25mm}|| }
\hline
\multicolumn{5}{|c|}{$E_{nl}$ $(E)$} \\
 \hline
$l$    & $E_{1l}$    & $E_{2l}$ & $E_{3l}$ & $E_{4l}$ \\
 \hline
$0$    &$none     $   &$none     $    &$none     $   &$none     $    \\
\hline
$1$    &$none     $   &$0.999988 $    &$none     $   &$none     $    \\ 
\hline
$2$    &$0.961220 $   &$none     $    &$0.999688 $   &$0.999988 $    \\
\hline
$3$    &$0.939873 $   &$0.958831 $    &$none     $   &$0.998511 $    \\
\hline
$4$    &$0.922010 $   &$0.937770 $    &$0.955919 $   &$none     $    \\
\hline
$5$    &$0.906141 $   &$0.919970 $    &$0.935276 $   &$0.952608 $    \\
\hline
$10$   &$0.841848 $   &$0.851190 $    &$0.860976 $   &$0.871257 $    \\
\hline
$50$   &$0.542193 $   &$0.544537 $    &$0.546944 $   &$0.549416 $    \\
\hline
$100$  &$0.308370 $   &$0.308232 $    &$0.308136 $   &$0.308080 $    \\
\hline
$500$  &$-0.488246$   &$-0.492937$    &$-0.497594$   &$-0.502218$    \\
\hline
$1000$ &$-0.783293$   &$-0.788018$    &$-0.792692$   &$-0.797315$    \\
\hline
$5000$ &$-0.989668$   &$-0.991032$    &$-0.992300$   &$-0.993473$    \\
\hline
$10000$&$-0.997896$   &$-0.998492$    &$-0.998990$   &$-0.999388$    \\
\hline
\end{tabular} 
\caption{Energy spectrum for the EMOS limit in three dimensions  for the higher values of $l$.}
\label{TabPSS3n}
\end{table}

\newpage
\begin{table}[ht]
    \centering
    \begin{tabular}{|p{10mm}||p{25mm}||p{25mm}||p{25mm}||p{25mm}|| }
\hline
\multicolumn{5}{|c|}{$E_{nl}$ $(E)$} \\
 \hline
$l$    & $E_{1l}$    & $E_{2l}$ & $E_{3l}$ & $E_{4l}$ \\
 \hline
$0$    &$none     $   &$none     $    &$none     $   &$none     $    \\
\hline
$1$    &$0.974544 $   &$0.999392 $    &$0.999988 $   &$none     $    \\ 
\hline
$2$    &$0.949943 $   &$0.971654 $    &$0.996735 $   &$0.999765 $    \\
\hline
$3$    &$0.930632 $   &$0.947749 $    &$0.968168 $   &$0.991779 $    \\
\hline
$4$    &$0.913876 $   &$0.928572 $    &$0.945107 $   &$0.964299 $    \\
\hline
$5$    &$0.898742 $   &$0.911841 $    &$0.926166 $   &$0.942099 $    \\
\hline
$10$   &$0.836270 $   &$0.845350 $    &$0.854842 $   &$0.864790 $    \\
\hline
$50$   &$0.539377 $   &$0.541684 $    &$0.544054 $   &$0.546487 $    \\
\hline
$100$  &$0.306397 $   &$0.306242 $    &$0.306128 $   &$0.306055 $    \\
\hline
$500$  &$-0.488753$   &$-0.493446$    &$-0.498104$   &$-0.502729$    \\
\hline
$1000$ &$-0.783454$   &$-0.788178$    &$-0.792852$   &$-0.797475$    \\
\hline
$5000$ &$-0.989670$   &$-0.991034$    &$-0.992302$   &$-0.993475$    \\
\hline
$10000$&$-0.997896$   &$-0.998492$    &$-0.998990$   &$-0.999388$    \\
\hline
\end{tabular} 
\caption{Energy spectrum for the EMOS limit in four dimensions for the higher values of $l$.}
\label{TabPSS4n}
\end{table}

\newpage
\begin{table}[ht]
    \centering
    \begin{tabular}{|p{10mm}||p{25mm}||p{25mm}||p{25mm}||p{25mm}|| }
\hline
\multicolumn{5}{|c|}{$E_{nl}$ $(E)$} \\
 \hline
$l$    & $E_{1l}$    & $E_{2l}$ & $E_{3l}$ & $E_{4l}$ \\
 \hline
$0$    &$none     $   &$0.999988 $    &$none     $   &$none     $    \\
\hline
$1$    &$0.961220 $   &$none     $    &$0.999688 $   &$0.999988 $    \\ 
\hline
$2$    &$0.939873 $   &$0.958831 $    &$none     $   &$0.998511 $    \\
\hline
$3$    &$0.922010 $   &$0.937770 $    &$0.955919 $   &$none     $    \\
\hline
$4$    &$0.906141 $   &$0.919970 $    &$0.935276 $   &$0.952608 $    \\
\hline
$5$    &$0.891630 $   &$0.904103 $    &$0.917617 $   &$0.932450 $    \\
\hline
$10$   &$0.830800 $   &$0.839636 $    &$0.848854 $   &$0.858495 $    \\
\hline
$50$   &$0.536574 $   &$0.538845 $    &$0.541178 $   &$0.543574 $    \\
\hline
$100$  &$0.304430 $   &$0.304258 $    &$0.304127 $   &$0.304036 $    \\
\hline
$500$  &$-0.489260$   &$-0.493954$    &$-0.498614$   &$-0.503240$    \\
\hline
$1000$ &$-0.783615$   &$-0.788339$    &$-0.793012$   &$-0.797634$    \\
\hline
$5000$ &$-0.989672$   &$-0.991036$    &$-0.992304$   &$-0.993477$    \\
\hline
$10000$&$-0.997896$   &$-0.998493$    &$-0.998990$   &$-0.999389$    \\
\hline
\end{tabular} 
\caption{Energy spectrum for the EMOS limit in five dimensions for the higher values of $l$.}
\label{TabPSS5n}
\end{table}

\newpage
\begin{table}[ht]
    \centering
    \begin{tabular}{|p{10mm}||p{25mm}||p{25mm}||p{25mm}||p{25mm}|| }
\hline
\multicolumn{4}{|c|}{$E_{10}$ $(E)$} \\
 \hline
$\delta$ $(E)$     & $D=3$$(dim)$    & $D=4$$(dim)$ & $D=5$$(dim)$ \\
 \hline
  & \multicolumn{3}{c||}{$a=1$ $(1/E)$} \\
\hline
$0.05$    &$none     $   &$-0.999738 $    &$-0.987390$       \\
\hline
          &$0.994216 $   &$0.993988  $    &$0.993568 $       \\ 
\hline
$0.10$    &$none     $   &$-0.998950 $    &$-0.948151$       \\
\hline
          &$0.982143 $   &$0.979277  $    &$0.973296 $      \\
\hline
  & \multicolumn{3}{c||}{$a=0$ $(1/E)$} \\
\hline
$0.05$    &$0.967789 $   &$0.964314 $    &$0.957414 $       \\
\hline
$0.10$    &$0.938819 $   &$0.915712 $    &$0.850753 $       \\
\hline
  & \multicolumn{3}{c||}{$a=-1$ $(1/E)$} \\
\hline
$0.05$    &$none     $   &$0.973259 $    &$none     $       \\
\hline
          &$none     $   &$0.775983 $    &$0.635179 $       \\ 
\hline
$0.10$    &$none     $   &$none     $    &$none     $       \\
\hline
          &$none     $   &$0.998412 $    &$0.417913 $      \\
\hline
\end{tabular} 
\caption{Ground state energy spectra in the pure vector limit.}
\label{TabPV}
\end{table}

\newpage
\begin{table}[ht]
    \centering
    \begin{tabular}{|p{10mm}||p{25mm}||p{25mm}||p{25mm}||p{25mm}|| }
\hline
\multicolumn{4}{|c|}{$E_{10}$ $(E)$} \\
 \hline
$\delta$ $(E)$     & $D=3$$(dim)$    & $D=4$$(dim)$ & $D=5$$(dim)$ \\
 \hline
  & \multicolumn{3}{c||}{$a=1$ $(1/E)$} \\
\hline
$0.05$    &$-0.903123$   &$-0.9765612$    &$none    $       \\
\hline
$0.10$    &$-0.731329$   &$-0.916126 $    &$none    $       \\
\hline
  & \multicolumn{3}{c||}{$a=0$ $(1/E)$} \\
\hline
$0.05$    &$none      $   &$none    $     &$none    $       \\
\hline
$0.10$    &$none      $   &$none    $     &$none    $       \\
\hline
  & \multicolumn{3}{c||}{$a=-1$ $(1/E)$} \\
\hline
$0.05$    &$0.903123  $   &$0.976512 $    &$none    $       \\
\hline
$0.10$    &$0.731329  $   &$0.916126 $    &$none    $       \\
\hline
\end{tabular} 
\caption{Ground state energy spectra in the pure scalar limit.}
\label{TabPS}
\end{table}

\newpage

\newpage

\newpage

\newpage

\newpage


\begin{thebibliography}{}

\bibitem{Schiff_Book} L.I. Schiff, \emph{Quantum Mechanics}, McGraw-Hill, New York, (1955).

\bibitem{Landau_Book} L.D Landau and E.M. Lifshitz, \emph{Quantum Mechanics, Non relativistic theory}, Pergamon, New York, (1977).

\bibitem{Flugge_Book} S. Fl\"ugge, \textit{Practical Quantum Mechanics}, Springer, Berlin, (1994).

\bibitem{Greiner_Book} W. Greiner, \emph{Relativistic Quantum Mechanics: Wave equations 3rd Ed.},  Springer, Berlin, (2000). 

\bibitem{Morse_1929} P.M. Morse, Phys. Rev. \textbf{34} (1929) 57.

\bibitem{Eckart_1930} C. Eckart, Phys. Rev. \textbf{35} (1930) 1303.

\bibitem{Rosen_Morse_1932} N. Rosen and P.M. Morse, Phys. Rev. \textbf{42} (1932) 210.

\bibitem{Manning_Rosen_1933} M.F. Manning and N. Rosen, Phys. Rev. \textbf{44} (1933) 953.

\bibitem{Poschl_Teller_1933} G. P\"oschl and E. Teller, Z. Physik \textbf{83} (1933) 143.

\bibitem{Yukawa_1935} H. Yukawa, Proc. Phys. Math. Soc. Jap.  \textbf{17} (1935) 48.

\bibitem{Hylleraas_1935} E.A. Hylleraas, J. Chem. Phys. \textbf{3} (1935) 595.

\bibitem{Hulthen_1942} L. Hulth\'en, Ark. Mat. Astron. Fys.  \textbf{28A} (1942) 5.


\bibitem{Woods_Saxon_1954} R.D. Woods and D.S. Saxon, Phys. Rev. \textbf{95} (1954) 577.

\bibitem{Klein_1926} O. Klein, Z. Phys. \textbf{37} (1926) 895.

\bibitem{Yi_et_al_2004} L.Z. Yi, Y.F. Diao, J.Y. Liu, and C.S. Jia, Phys. Lett. A. \textbf{333} (2004) 212.

\bibitem{Villalba_et_al_2006} V.M. Villalba and C. Rojas, Int. J. Mod. Phys. A \textbf{21} (2006) 313.

\bibitem{Olgar_et_al_2006} E. Olgar, R. Koc, and H. Tutunculer, Chin. Phys. Lett. \textbf{23} (2006) 539.

\bibitem{Ciftci_et_al_2003} H. Ciftci, R.L. Hall, and N. Saad, J. Phys. A: Math. Gen. \textbf{36} (2003) 11807.

\bibitem{Olgar_et_al_2008} E. Ol\u{g}ar, R. Koc, and H. Tutunculer, Phys. Scr. \textbf{78} (2008) 015011.

\bibitem{Olgar_2008} E. Ol\u{g}ar, Chin. Phys. Lett. \textbf{25} (2008) 1939.

\bibitem{Xu_et_al_2010} Y. Xu, S. He, and C.S. Jia, Phys. Scr. \textbf{81} (2010) 045001.

\bibitem{Ikot_et_al_2012} A.N. Ikot, O.A. Awoga, and B.I. Ita, Few-Body Syst. \textbf{53} (2012) 539.

\bibitem{Jia_et_al_2013} C.S. Jia, T. Chen, and  S. He, Phys. Lett. A \textbf{377} (2013) 682.

\bibitem{Hou_et_al_1999} C.F. Hou, Z.X. Zhou, and Y. Li, Acta Phys. Sin. (Overseas Edn) \textbf{8} (1999) 561.

\bibitem{Rojas_et_al_2005} C. Rojas and V.M. Villalba, Phys. Rev. A \textbf{71} (2005) 052101.

\bibitem{Hassanabadi_et_al_2013} H. Hassanabadi, E. Maghsoodi, S. Zarrinkamar, and N. Salahi, Few-Body Syst. \textbf{54} (2013) 2009. 

\bibitem{Arda_et_al_2009} A. Arda and R. Sever, Int. J. Mod. Phys. A \textbf{24} (2009) 3985.

\bibitem{Badalov_et_al_2010} V.H. Badalov, H.I. Almadov, and S.V. Badalov, Int. J. Mod. Phys. E \textbf{19} (2010) 1463.

\bibitem{Bayrak_et_al_2015D} O. Bayrak and D. Sahin, Commun. Theor. Phys. \textbf{64} (2015) 259.

\bibitem{Bayrak_et_al_2015E} O. Bayrak and E. Aciksoz, Phys. Scr. \textbf{90} (2015) 015302.

\bibitem{Lutfuoglu_et_al_2018_LK} B.C.  L\"{u}tf\"{u}o\u{g}lu, J. Lipovsk\'y, and J. K\v{r}\'{i}\v{z}, Eur. Phys. J. Plus \textbf{133} (2018) 17. 

\bibitem{Lutfuoglu_et_al_2018} B.C.  L\"{u}tf\"{u}o\u{g}lu, Eur. Phys. J. Plus \textbf{133} (2018) 309.

\bibitem{Diao_et_al_2004} Y.F. Diao, L.Z. Yi, and C.S. Jia, Phys. Lett. A \textbf{332} (2004) 157.

\bibitem{Olgar_2009} E. Ol\u{g}ar, Chin. Phys. Lett. \textbf{26} (2009) 020302.

\bibitem{Lutfuoglu_Ikot_et_al_2018} B.C. L\"utf\"uo\u glu, A.N. Ikot, E.O. Chukwocha, and F.E. Bazuaye, Eur. Phys. J. Plus \textbf{133} (2018) 528.

\bibitem{Chen_et_al_2008} C.Y. Chen, F.L. Lu, and D.S. Sun, Phys. Scr. \textbf{78} (2008) 015014.

\bibitem{Ortakaya_2012} S. Ortakaya, Chin. Phys. B. \textbf{21} (2012) 070303.


\bibitem{Dong_Book} S.H. Dong, \textit{Factorization Method in Quantum Mechanics}, Springer, Netherland, (2007).

\bibitem{Chen_et_al_2003} C.Y. Chen, C.L. Liu, F.L. Lu, and, D.S. Sun, Acta Phys. Sinica \textbf{52} (2003) 1579.

\bibitem{Saad_et_al_2008} N. Saad, R.L. Hall, and H. Ciftci, Cent. Eur. J. Phys. \textbf{6} (2008) 717.

\bibitem{Hassanabadi_et_al_2011} H. Hassanabadi, H. Rahimov, and S. Zarrinkamar, Adv. High Energy Phys. \textbf{2011} (2011) 458087.

\bibitem{Hassanabadi_et_al_2012} H. Hassanabadi, E. Maghsoodi, S. Zarrinkamar, and H. Rahimov, Eur. Phys. J. Plus \textbf{127} (2012) 143.

\bibitem{Ibrahim_et_al_2012} T.T. Ibrahim, K.J. Oyewumi, and S.M. Wyngaardt, Eur. Phys. J. Plus \textbf{127} (2012) 100.

\bibitem{Ortakaya_2013} S. Ortakaya, Chin. Phys. B. \textbf{22} (2013) 070303.

\bibitem{Antia_et_al_2013} A.D. Antia, A.N. Ikot, H. Hassanabadi, and E. Maghsoodi, Ind. J. Phys. \textbf{87} (2013) 1133.

\bibitem{Chen_et_al_2014} X.Y. Chen, T. Chen, and C.S. Jia, Eur. Phys. J. Plus \textbf{129} (2014) 75.

\bibitem{Ikot_et_al_2014} A.N. Ikot, H. Hassanabadi, H.P. Obong, Y.E. Chad-Umoren, C.N. Isonguyo, and B.H. Yazarloo, Chin. Phys. B \textbf{23} (2014) 120303.

\bibitem{Tan_et_al_2014} M.S. Tan, S. He, and C.S. Jia, Eur. Phys. J. Plus \textbf{129} (2014) 264.

\bibitem{Jia_et_al_2015} C.S. Jia, J.W. Dai, L.H. Zhang, J.Y. Liu, and G.D. Zhang, Chem. Phys. Lett. \textbf{619} (2015) 54.

\bibitem{Xie_et_al_2015} X.J. Xie and C.S. Jia, Phys. Scr. \textbf{90} (2015) 035207.

\bibitem{Ikot_Lutfuoglu_2016} A.N. Ikot, B.C. L\"utf\"uo\u{g}lu, M.I. Ngwueke, M.E. Udoh, S. Zare, and H. Hassanabadi, Eur. Phys. J. Plus \textbf{131} (2016) 419.


\bibitem{Ikhdair_et_al_2013} S.M. Ikhdair and B.J. Falaye, Chem. Phys. \textbf{421} (2013) 84.

\bibitem{Oyewumi_et_al_2014} K.J. Oyewumi, B.J. Falaye, C.A. Onate, O.J. Oluwadare, and W.A. Yahya, Mol. Phys. \textbf{112} (2014) 127.

\bibitem{Onate_et_al_2015} C.A. Onate and J.O. Ojonubah, Int. J. Mod. Phys. E \textbf{24} (2015) 1550020.

\bibitem{Arda_et_al_2016} A. Arda, C. Tezcan, and R. Sever, Few-Body Sys. \textbf{57} (2016) 93. 

\bibitem{Onyeaju_et_al_2017} M.C. Onyeaju, A.N. Ikot, C.A Onate, O. Ebomwonyi, M.E. Udoh, and J.O.A. Idiodi, Eur. Phys. J. Plus \textbf{132} (2017) 302.

\bibitem{Ikot_et_al_2018} A.N. Ikot, E.O. Chukwuocha, M.C. Onyeaju, C.A. Onate, B.I. Ita, and M.E. Udoh, Pramana J. Phys. \textbf{90} (2018) 22.

\bibitem{Ortega_et_al_2018} G. Valencia-Ortega and L.A. Arias-Hernandex, Int. J. Quant. Chem. \textbf{118} (2018) e25589. 

\bibitem{Okorie_et_al_2018_a} U.S. Okorie, A.N. Ikot, M.C. Onyeaju, and E.O. Chukwuocha, J. Mol. Model. \textbf{24} (2018) 289. 

\bibitem{Okorie_et_al_2018_b} U.S. Okorie, E.E. Ibekwe, A.N. Ikot, M.C. Onyeaju, and E.O. Chukwuocha, J. Kor. Phys. Soc. \textbf{73} (2018) 1211.

\bibitem{Okorie_et_al_2018_c} U.S. Okorie, A.N. Ikot, M.C. Onyeaju, and  E.O. Chukwuocha, Rev. Mex. Fis. \textbf{64} (2018) 608. 

\bibitem{Okorie_et_al_2018_d} U.S. Okorie, E.E. Ibekwe,  M.C. Onyeaju, and A.N. Ikot, Eur. Phys. J. Plus \textbf{133} (2018) 433.

\bibitem{Ikot_et_al_2019_e} A.N. Ikot, W. Azogor, U.S. Okorie, F.E. Bazuaye,  M.C. Onyeaju, C.A. Onate and E.O. Chukwuocha, Ind. J. Phys. \textbf{} (2019) .... . https://doi.org/10.1007/s12648-019-01375-0

\bibitem{Ikot_et_al_2019_ML} A.N. Ikot, U.S. Okorie, C.A. Onate, M.C. Onyeaju, and H. Hassanabadi, Can. J. Phys. \textbf{} (2019) .... .  https://doi.org/10.1139/cjp-2018-0535

\bibitem{Lutfuoglu_et_al_2016} B.C.  L\"{u}tf\"{u}o\u{g}lu, F. Akdeniz, and O. Bayrak, J. Math. Phys. \textbf{57} (2016) 032106.

\bibitem{Lutfuoglu_2019} B.C.  L\"{u}tf\"{u}o\u{g}lu, Common. Theor. Phys. \textbf{71} (2019) 267.

\bibitem{Lutfuoglu_et_al_2016_ME} B.C. L\"utf\"uo\u{g}lu and M. Erdogan, Anadolu Uni. J. Sci. Tech. A Appl. Sci. Eng. \textbf{17} (2016) 708.

\bibitem{Lutfuoglu_et_al_2017_ME} B.C. L\"utf\"uo\u{g}lu and M. Erdogan, S\"uleyman Demirel Uni. J. Nat. Appl. Sci. \textbf{21} (2017) 316.

\bibitem{Lutfuoglu_2018_1} B.C.  L\"{u}tf\"{u}o\u{g}lu, Common. Theor. Phys. \textbf{69} (2018) 23.

\bibitem{Lutfuoglu_2018_2} B.C.  L\"{u}tf\"{u}o\u{g}lu, Can. J. Phys. \textbf{96} (2018) 853.

\bibitem{Lutfuoglu_et_al_2019} B.C.  L\"{u}tf\"{u}o\u{g}lu and J. K\v{r}\'{i}\v{z}, Eur. Phys. J. Plus \textbf{134} (2019) 60.

\bibitem{Chen_et_al_2013} T. Chen, S.R. Lin, and C.S. Jia, Eur. Phys. J. Plus \textbf{128} (2013) 69.

\bibitem{Hu_et_al_2014} X.T Hu and C.S. Jia, Can. J. Chem. \textbf{92} (2014) 386.

\bibitem{Jia_et_al_2017_1} C.S. Jia, L.H. Zhang, and C.W. Wang, Chem. Phys. Lett. \textbf{667} (2017) 211.

\bibitem{Song_et_al_2017_2} X.Q. Song, C.W. Wang, and C.S. Jia, Chem. Phys. Lett. \textbf{673} (2017) 50.

\bibitem{Jia_et_al_2017_3} C.S. Jia, C.W. Wang, L.H. Zhang, X.L. Peng, R. Zeng, and X.T. You, Chem. Phys. Lett. \textbf{676} (2017) 150.

\bibitem{Wang_et_al_2017_4} J.F. Wang, X.L. Peng, L.H. Zhang, C.W. Wang, and C.S. Jia, Chem. Phys. Lett. \textbf{686} (2017) 131.

\bibitem{Jia_et_al_2018_1} C.S. Jia, C.W. Wang, L.H. Zhang, X.L. Peng, H.M. Tang, and R. Zeng, Chem. Eng. Sci. \textbf{183} (2018) 26.

\bibitem{Jia_et_al_2018_2} C.S. Jia, R. Zeng, X.L. Peng, L.H. Zhang, and Y.L. Zhao, Chem. Eng. Sci. \textbf{190} (2018) 1.

\bibitem{Jia_et_al_2018_3} X.L. Peng, R. Jiang, C.S. Jia,  L.H. Zhang,  and Y.L. Zhao , Chem. Eng. Sci. \textbf{190} (2018) 122.

\bibitem{Ocak_et_al_2018} Z. Ocak, H. Yanar, M. Salti, and O. Aydogdu, Chem. Phys. \textbf{513} (2018) 252.

\bibitem{Deng_et_al_2018} M. Deng and C.S. Jia, Eur. Phys. J. Plus \textbf{133} (2018) 25.

\bibitem{Jia_et_al_2018_4} C.S. Jia, C.W. Wang, L.H. Zhang, X.L. Peng, H.M. Tang, J.Y. Liu, Y. Xiong, and R.Zeng, Chem. Phys. Lett. \textbf{692} (2018) 57.

\bibitem{Khordad_et_al_2019} R. Khordad, A. Avazpour, and A. Ghanbari, Chem. Phys. \textbf{517} (2019) 30.

\bibitem{Jia_et_al_2019_ek_1} C.S. Jia, L.H. Zhang, X.L Peng, J.X. Luo, Y.L. Zhao, J.Y. Liu, J.J. Guo, and L.D. Tang, Chem. Eng. Sci. \textbf{202} (2019) 70. 

\bibitem{Jiang_et_al_ek_2} R. Jiang, C.S. Jia, Y.Q. Wang, X.L. Peng, and 
L.H. Zhang, Chem. Phys. Lett. \textbf{715} (2019) 186.

\bibitem{Jia_et_al_2019_ek_3} C.S. Jia, X.T. You, J.Y. Liu, L.H. Zhang, X.L. Peng, Y.T. Wang, and L.S. Wei, Chem. Phys. Lett. \textbf{717} (2019) 16.

\bibitem{Jiang_et_al_ek_4} R. Jiang, C.S. Jia, Y.Q. Wang, X.L. Peng, and 
L.H. Zhang, Chem. Phys. Lett. \textbf{726} (2019) 83.



\bibitem{Synder_1940} H. Snyder and J. Weinberg, Phys. Rev. \textbf{57} (1940) 307.

\bibitem{Schiff_1940} L.I. Schiff, H. Snyder, and J. Weinberg, Phys. Rev. \textbf{57} (1940) 315.

\bibitem{Green_1962} A.M. Green, Nucl. Phys. \textbf{33} (1962) 218.

\bibitem{Formanek_et_al_2004} J. Formanek, R.J. Lombard, and J. Mares, Czech J. Phys. \textbf{54} (2004) 289.

\bibitem{Lombard_et_al_2007} R.J. Lombard, J. Mares, and C. Volpe, J. Phys. G: Nucl. Part. Phys. \textbf{34} (2007) 1879.

\bibitem{Benchikha_et_al_2013}  A. Benchikha and L. Chetouani, Mod. Phys. Lett. A \textbf{28}  (2013) 1350079.

\bibitem{Benchikha_et_al_2014}  A. Benchikha and L. Chetouani, Cent. Eur. J. Phys. \textbf{12}  (2014) 392.

\bibitem{Gupta_et_al_2012} P. Gupta and I. Mehrotra, J. Mod. Phys. \textbf{3} (2012) 1530.

\bibitem{Ikot_Hassanabadi_2013} A.N. Ikot, H. Hassanabadi, E. Maghsoodi, and S. Zarrinkamer, Ukr. J. Phys. \textbf{58} (2013) 915.

\bibitem{Boumali_et_al_2017} A. Boumali, S. Dilmi, S. Zare, and H. Hassanabadi, KIJOMS \textbf{3} (2017) 191.

\bibitem{Boumali_et_al_2018} A. Boumali and M. Labidi, Mod. Phys. Lett. A \textbf{33} (2018) 1850033.



\bibitem{Greene_et_al_1976} R.L. Greene and C. Aldrich, Phys. Rev. A \textbf{14} (1976) 2363.

\bibitem{Abramowitz_et_al_Book} M. Abramowitz and I.A. Stegun, \textit{Handbook of Mathematical Functions with formulas}, Dover, New York (1972).



\end{thebibliography}
\end{document}